# Orthoexciton-to-paraexciton conversion rate in $Cu_2O$ at sub-Kelvin temperatures

Yusuke Morita[1,2,3], Yukihiro Akiyama[1], Yuki Haita[1], Yuga Hirota[1] and Kosuke Yoshioka[1,4,†]

[1] *Department of Applied Physics, Graduate School of Engineering, The University of Tokyo, 2-11-16 Yayoi, Bunkyo-ku, Tokyo, 113-8656, Japan*
[2] *Institute for Photon Science and Technology, The University of Tokyo, 7-3-1 Hongo, Bunkyo-ku, Tokyo, 113-0033, Japan*
[3] *RIKEN Center for Emergent Matter Science (CEMS), 2-1 Hirosawa, Wako, Saitama 351-0198, Japan*
[4] *Photon Science Center, Graduate School of Engineering, The University of Tokyo, 2-11-16 Yayoi, Bunkyo-ku, Tokyo 113-0032, Japan*

We measured the conversion rate from 1*s* orthoexcitons to 1*s* paraexcitons in $Cu_2O$ below 1 K using a dilution refrigerator. To determine the conversion rate, we performed time-resolved luminescence spectroscopy of 1*s* orthoexcitons and 1*s* paraexcitons confined in a strain-induced trap potential. From the time evolution of the $\Gamma_3^-$-LO-phonon-assisted luminescence of 1*s* orthoexcitons, we obtained a conversion rate of $5.9 \times 10^7$ s$^{-1}$ at a mixing-chamber temperature of 64 mK under an applied stress of 2.2 kbar. Between 64 and 800 mK, the conversion rate depends only weakly on temperature, and it is less than half of the value of $1.6 \times 10^8$ s$^{-1}$ reported at 2 K under the same applied stress. The conversion rate therefore continues to decrease below 2 K, contrary to the previous expectation that it becomes nearly constant below about 1.7 K. We discuss these results in terms of the momentum distribution of orthoexcitons, which does not reach equilibrium with the lattice during the conversion at these low temperatures. These findings call for a theory that accounts for both the small magnitude of the conversion rate and its weak temperature dependence below 1 K.

## I. INTRODUCTION

Bulk $Cu_2O$ has long been a textbook system for studying properties of excitons in semiconductors [1, 2, 3]. Early spectroscopic and theoretical studies identified wavefunctions of the yellow series exciton states, which consist of an electron in the lowest conduction band ($\Gamma_6^+$) and a hole in the highest valence band ($\Gamma_7^+$) [4]. The lowest exciton state in the yellow series, which has a principal quantum number $n$=1, has a large binding energy of 150 meV. The exciton state is split into a triply degenerate orthoexciton state and a nondegenerate paraexciton state, depending on the spin configuration of an electron and a hole. The energy of the 1*s* orthoexciton state is 12 meV higher than the energy of the 1*s* paraexciton state. While 1*s* orthoexcitons are optically accessible, 1*s* paraexcitons are optically forbidden to all orders of direct transitions by spin and parity selection rules under no external field. The weak coupling between 1*s* paraexcitons and the radiation field results in an extremely long lifetime of 1*s* paraexcitons [5, 6]. The spin conversion of photoexcited 1*s* orthoexcitons to 1*s* paraexcitons is a fundamental process for the generation of long-lived 1*s* paraexcitons.

Spin conversion and spin relaxation processes of excitons have been studied in a variety of systems, including bulk crystals [7, 8], quantum wells [9, 10], perovskites [11, 12], and atomically thin transition-metal dichalcogenides [13]. Several theoretical models have also been proposed to explain spin relaxation of excitons, such as the D’yakonov–Perel’ mechanism, the Elliott–Yafet mechanism, and the mechanism based on the long-range electron-hole exchange interaction [14]. Ref. [14] noted that the Elliott–Yafet mechanism and the D’yakonov–Perel’ mechanism can contribute to spin relaxation of exciton in bulk GaAs. In GaAs quantum wells, the spin relaxation of excitons has been explained mainly by the long-range electron-hole exchange interaction [14, 15], although the spin flip of an individual electron or hole can also contribute [14]. In contrast, excitons in bulk $Cu_2O$ have unique features: the crystal has inversion symmetry, which rules out the D’yakonov–Perel’ mechanism, and the energy splitting between 1*s* orthoexcitons and 1*s* paraexcitons originates from the short-range electron-hole exchange interaction. Therefore, an investigation of the spin conversion from 1*s* orthoexciton to 1*s* paraexciton in bulk $Cu_2O$ is important for studying the spin conversion mechanism of excitons.

The conversion process from 1*s* orthoexcitons to 1*s* paraexcitons in $Cu_2O$ has been extensively studied over several decades. Kreingold and Makarov experimentally investigated the conversion process for the first time [16]. They measured luminescence intensities of 1*s* orthoexcitons and 1*s* paraexcitons as a function of the lattice temperature in the range of 4.2–40 K. They observed that the intensity of the 1*s* orthoexciton luminescence increases exponentially above 20 K and exhibits a minimum at a temperature of about 20 K. The exponential temperature dependence above 20 K suggested that 1*s* orthoexcitons and 1*s* paraexcitons are in thermal equilibrium. In contrast, the minimum in the 1*s* orthoexciton luminescence intensity around 20 K implies that thermal equilibrium between 1*s* orthoexcitons and 1*s* paraexcitons does not hold below 20 K. From these results, they concluded that the conversion rate is temperature dependent and that large conversion rates above 20 K enable thermal equilibrium. Mysyrowicz *et al.* discussed the mechanism of the conversion process based on measurements of luminescence intensities of 1*s* orthoexcitons and 1*s* paraexcitons over the temperature range 1.5–30 K [17]. Weiner *et al.* measured the lifetime of 1*s* orthoexcitons in the range of 2.5–30 K with subnanosecond time resolution [18]. They also obtained evidence that the lifetime of 1*s* orthoexcitons is determined by the conversion to 1*s* paraexcitons. Accordingly, they evaluated the conversion rate as a function of the lattice temperature. The purple triangles in Fig. 1 show the results. They found that the decay rate is well described by $\alpha + \beta T^{3/2}$ where $\alpha = 4\times10^{8}\ \mathrm{s}^{-1}$ and $\beta = 0.3\times10^{8}\ \mathrm{s}^{-1}\cdot\mathrm{K}^{-3/2}$. Jang *et al.* measured the conversion rate over the temperature range 1.7–20 K using time-resolved photoluminescence spectroscopy [19]. Green squares in Fig. 1 show the results. They estimated the low temperature limit of the conversion rate to be $3.0\times10^{8}\ \mathrm{s}^{-1}$.

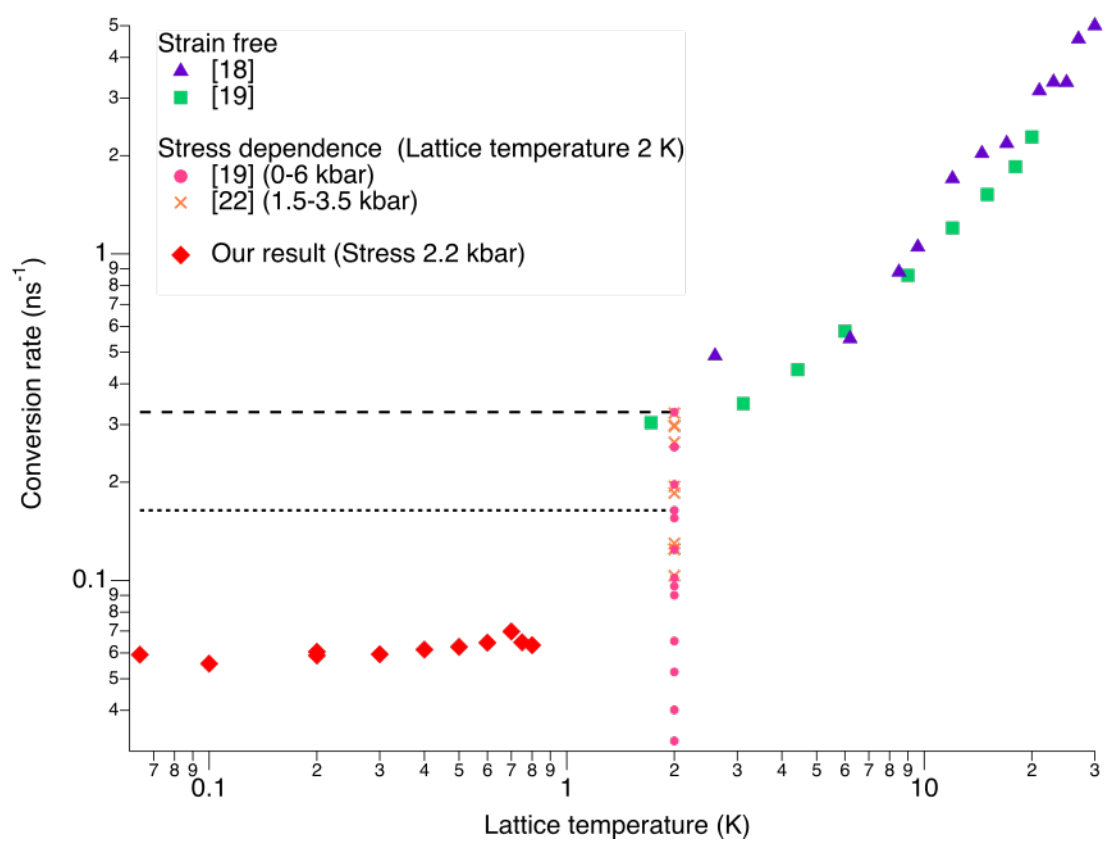

FIG. 1. Lattice temperature dependence of the previously reported orthoexciton-paraexciton conversion rates measured above 1.7 K and rates at sub-Kelvin temperatures measured in the present work. In the present experiments, the lattice temperature corresponds to the mixing-chamber temperature. The purple triangles (green squares) are reported in Ref. [18] (Ref. [19]). These results were obtained under strain-free conditions in the crystal. Orange crosses (pink circles) points show the conversion rate at a lattice temperature of 2 K under applied stress in the range 1.5-3.5 kbar (0-6 kbar), reported in Ref. [22] (Ref. [19]). The dashed line shows the conversion rate of $3.0\times10^{8}$ s$^{-1}$ at 2 K under a stress-free condition, reported in Ref. [19]. The dotted line shows the conversion rate of $1.6\times10^{8}$ s$^{-1}$ at 2 K under 2.2 kbar, which is the same stress condition as in our experiment. Red diamonds show our results at 2.2 kbar.

The mechanism of the conversion process, which accounts for the conversion rate above 1.7 K, has been studied. Caswell and Yu [20] proposed a two-phonon emission model to explain the temperature dependence of the intensity of 1*s* orthoexciton luminescence reported in Ref. [16]. In this model, 1*s* orthoexcitons convert to 1*s* paraexcitons through the emission of an optical phonon and a longitudinal acoustic (LA) phonon. However, this model was ruled out by the stress dependence of the conversion rate [21]. Furthermore, the stress dependence of the conversion rate at a lattice temperature of 2 K was reported in Ref. [22]. The stress applied via the Hertzian contact ranged from 1.5 to 3.5 kbar. The orange crosses in Fig. 1 show conversion rates at these stresses. The conversion rate at 3.5 kbar is $1\times10^{8}$ s$^{-1}$, which is approximately one-third of the value of $3\times10^{8}$ s$^{-1}$ at 1.5 kbar. This gradual decrease supports the conclusion of Ref. [21]. The single-acoustic-phonon emission model proposed in Ref. [21] and refined in Ref. [22] can explain the stress dependence of the conversion rate. Furthermore, Jang *et al.* also measured the conversion rate at a lattice temperature of 2 K as a function of the stress in the range 0–6 kbar [19]. The results (shown by the pink circles in Fig. 1) are in good agreement with those of Ref. [22]. Reference [19] proposes a mechanism in which the emission or absorption of a single transverse acoustic (TA) phonon causes the spin flip through the rotation of the lattice. This model explains both the stress and temperature dependences of the conversion rate above 1.7 K, although the calculated rates are about 150 times smaller than the experimental rates. This model predicts that the conversion rate converges to a constant value below 1.7 K.

Previous studies have established a widely accepted qualitative picture of the acoustic-phonon-mediated conversion process above 1.7 K, while the conversion process at lower temperatures remains unclear. As shown in Fig. 1, the conversion rate appears to converge to about $3\times 10^{8}\ s^{-1}$ at 1.7 K and exhibits a power-law dependence on the lattice temperature at high temperatures. Application of stress to the crystal leads to a gradual decrease in the conversion rate. Below 1.7 K, the conversion rate has not been determined quantitatively. In our previous study at sub-Kelvin temperatures, the luminescence of 1*s* paraexcitons appeared clearly only in the time gate following that of the orthoexciton luminescence, with a gate width of 30 ns [23]. This result suggested that the conversion is slower than at 1.7 K, but the time resolution was not sufficient to determine the conversion rate. Measurements in this temperature range are required to confirm whether the conversion rate at 1.7 K approaches its low-temperature limit. Measurements of the temperature dependence at lower temperatures also help elucidate the mechanism of the conversion process. In addition, understanding the conversion process at sub-Kelvin temperatures would be essential for investigating the formation mechanism of the Bose–Einstein condensate (BEC) of 1*s* paraexcitons recently reported at sub-Kelvin temperatures [24]. Conversion rates measured at about 2 K have also been used as input parameters in the analysis of exciton dynamics at sub-Kelvin temperatures [25], although this assumption has not been tested.

†Contact author: yoshioka@fs.t.u-tokyo.ac.jp

In this work, we report measurements of the conversion rate below 1 K using time-resolved photoluminescence spectroscopy. We used a cryogen-free dilution refrigerator to cool a $Cu_2O$ crystal to 64 mK [24, 26]. We applied an inhomogeneous stress to the crystal at 2.2 kbar via the Hertzian contact. The inhomogeneous stress results in an inhomogeneous strain field that acts as a trap potential for both 1*s* orthoexcitons and 1*s* paraexcitons. The strain-induced trap potential with a slightly different applied stress was implemented in Ref. [24], which reports the observation of the exciton BEC. Furthermore, the application of stress facilitates good thermal contact between 1*s* orthoexcitons and the lattice at sub-Kelvin temperatures [26]. The thermal contact reduces the discrepancy of the temperature between 1*s* orthoexcitons and the lattice, as required for the present experiment. A short excitation pulse generated 1*s* orthoexcitons near the bottom of the trap potential. We observed time-resolved $\Gamma^{-}_{3}$–longitudinal-optical (LO)-phonon-assisted luminescence from 1*s* orthoexcitons in the trap potential using a gated intensified charge-coupled device (ICCD) camera. From the temporal evolution of the $\Gamma_3$–phonon-assisted luminescence, as discussed in detail below, we obtained conversion rates smaller than that those expected from the previously known temperature dependence above 1.7 K.

## II. METHODS

We used a naturally grown pure single crystal of $Cu_2O$. The sample is identical to that in Ref. [24]. The sample surface consisted of the (100), (011), and $(01\bar{1})$ crystal surface. The sample dimensions were 5.3 × 5.3 × 8.0 $mm^3$. The sample was mounted in an apparatus based on a cryogen-free dilution refrigerator (Oxford Instruments, DR-400), as described in Ref. [24]. The lowest base temperature was 64 mK. We controlled the base temperature in the range from 64 to 800 mK using a heater attached to the mixing chamber in the dilution refrigerator.

We applied inhomogeneous stress via the Hertzian contact between the (100) crystal surface and the spherical surface of a lens (BK7, radius of curvature = 15.57 mm). The applied stress was estimated to be 2.2 kbar. The stress results in an inhomogeneous strain field. This strain field leads to inhomogeneous energy shifts in the 1*s* orthoexcitons and the 1*s* paraexcitons. The energy shifts form three-dimensional trap potentials for each. The relation between the applied stress via Hertzian contact and the energy shifts of the 1*s* orthoexcitons and the 1*s* paraexcitons has been well established [27, 28]. Based on these studies, we calculated the spatial distribution of the energy shifts of 1*s* orthoexcitons and 1s paraexcitons. Fig. 2a shows the calculated energy shifts along the (011) crystal axis at a distance of 285 μm from the (100) crystal surface, which corresponds to the bottom of the trap potential for 1*s* orthoexcitons.

We generated 1*s* orthoexcitons via the $\Gamma^{-}_{3}$-LO-phonon-assisted absorption process, using an excitation pulse with a duration of approximately 100 picoseconds. The resonant wavelength of the absorption process depends on the position in the trap potential, because the 1*s* orthoexciton energy level varies with the strain distribution. We used the excitation light with a narrow spectral width spanning from 607.4 to 607.5 nm, as described below, to generate 1*s* orthoexcitons near the bottom of the trap potential. The conversion process of 1*s* orthoexcitons is the dominant decay process, compared with the radiative decay processes [29]. As a result, almost all 1*s* orthoexcitons are converted to 1*s* paraexcitons. The 1*s* paraexcitons have a lifetime of approximately 400 ns as reported in Ref. [6].

We employed a Ti:sapphire femtosecond regenerative amplifier (Coherent Inc., Libra HE) as the excitation light source. The fundamental pulse was centered at 800 nm with a repetition rate of 1 kHz, and the pulse energy was 2.8 mJ. Figure 2(b) illustrates the optical setup used to prepare the excitation light. The

† Contact author: yoshioka@fs.t.u-tokyo.ac.jp

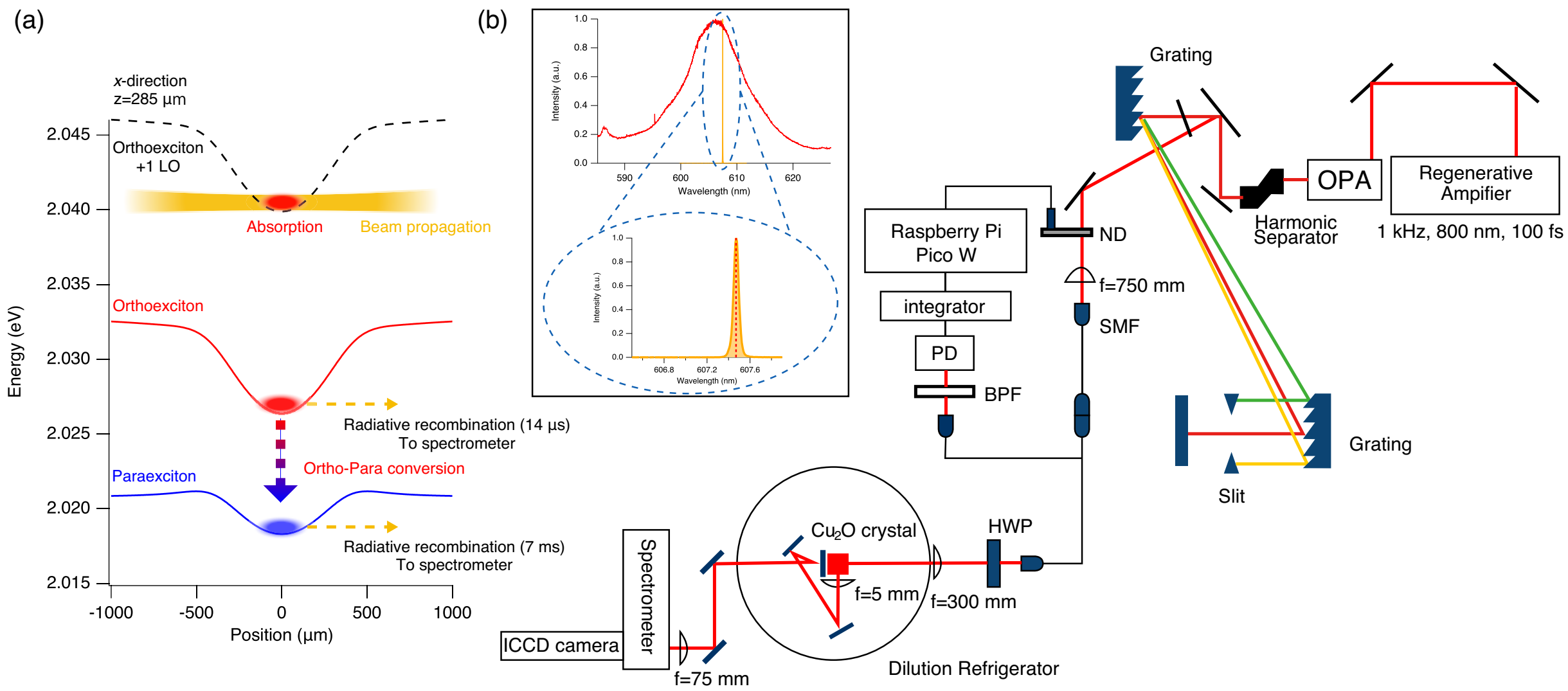

FIG. 2 (a) Schematic illustration of the optical excitation and spectroscopy of 1*s* orthoexcitons and 1*s* paraexcitons. The dashed curve shows the position-dependent resonant photon energy associated with the $\Gamma^{-}_{3}$-LO- phonon-assisted absorption of 1*s* orthoexcitons along the [011] crystal axis in the trap potential. The red (blue) curve shows the spatial distribution of the potential energy for 1*s* orthoexcitons (1*s* paraexcitons). The yellow shaded region indicates the propagation direction of the excitation beam as well as its spectral range. The red shaded region represents the spatial extent of 1*s* orthoexcitons generated via the phonon-assisted absorption process. We observed their radiative recombination. We also observed the radiative recombination of 1*s* paraexcitons.
(b) Experimental setup with the optical path for the excitation light and the optical path for time-resolved luminescence spectroscopy. OPA, ND, SMF, PD, BPF, HWP, and ICCD Camera denote optical parametric amplifier, neutral density filter, single-mode fiber, photo detector, bandpass filter, half-wave plate, and intensified charge-coupled device (ICCD) camera, respectively. We used an integrator circuit and a Raspberry Pi Pico W to construct a feedback system for stabilizing the excitation power.
(Inset) Spectra of the sum-frequency light emitted from the optical parametric amplifier (red curve) and the spectrally filtered light (yellow curve). The central wavelength of the filtered light was 607.47 nm, which corresponds to the resonant wavelength of the $\Gamma^{-}_{3}$-LO- phonon-assisted absorption of 1*s* orthoexcitons at the bottom of the trap potential.

fundamental pulse was used to pump a femtosecond optical parametric amplifier (OPA) (Light Conversion Inc., TOPAS-C). The OPA generates the second harmonic of the signal. The second-harmonic light had a broad spectral width spanning from 590 to 620 nm. The second-harmonic light was directed into an external spectral filter, which consisted of a pulse stretcher configuration incorporating double diffraction gratings and a slit positioned in front of one grating. The spectrally filtered light was centered at 607.47 nm with a narrow spectral bandwidth of 0.1 nm (full width at half maximum, FWHM). The inset in Fig. 2b shows the spectra of the second-harmonic light and the spectrally filtered light. The spectrally filtered light was coupled into a 5 m-long single-mode fiber. The estimated pulse duration was 100 ps, was 100 ps. Considering the long duration and the pulse energy employed in this study, the spectral broadening by self-phase modulation was negligible. We also constructed a feedback system to control the transmittance of a neutral density (ND) filter so that the average excitation power was stabilized at 300 nW. Our measurements typically required long data acquisition times of around 3 hours. Therefore, stabilization of the excitation light power is essential. The excitation beam was introduced to the sample along the (011) crystal axis. A focusing lens with a focal length of 300 mm set outside the refrigerator focused the excitation beam on the bottom of the trap potential. The spot size (FWHM) of the focused excitation light was 30 µm.

We observed the space- and time-resolved $\Gamma^{-}_{3}$–LO-phonon-assisted luminescence spectrum of 1*s* orthoexcitons. We also measured the space- and time-resolved luminescence spectra of 1*s* orthoexcitons and 1*s* paraexcitons via the direct recombination process.

† Contact author: yoshioka@fs.t.u-tokyo.ac.jp

To collect luminescence from 1*s* orthoexcitons and 1*s* paraexcitons with a large solid angle, a ZnSe meniscus lens (focal length: 5 mm, effective aperture: 6 mm) was employed as an objective lens positioned in close proximity to the $Cu_2O$ crystal. The luminescence images were magnified by a factor of 15. We employed a 50-cm imaging spectrometer (Acton Research Corp., Spectra Pro-500i) to resolve the photon energies of the luminescence. The entrance slit of the spectrometer was set to its maximum width of 3.1 mm to increase the detection efficiency of the luminescence images. The energy resolution was 600 μeV. We used an ICCD camera (Andor Technology, iStar DH334T). The luminescence signal was integrated while an electrical gate pulse was applied to the ICCD camera. The electrical gate pulse was generated by an internal gate generator of the camera. We used an electrical pulse from the Synchronization and Delay Generator (SDG) unit, which controls the timing of the Pockels cell in the regenerative amplifier, as an external trigger for the gate generator. The delay of the gate pulse with respect to the excitation pulse was controlled through the interval between the gate pulse and the external trigger pulse. By varying this delay, we obtained the time evolution of the spatially resolved luminescence spectra. The central region for the observation of 1*s* orthoexcitons was positioned at the bottom of the trap potential. The dimensions of the observation region were 206 μm along the [011] axis, 800 μm along the [100] axis, and 30 μm along the $[01\bar{1}]$ axis. Three dimensions were determined by the slit width, the size of the two-dimensional CCD camera in the slit direction, the image magnification, and the depth of focus of the imaging system. 1*s* orthoexcitons were selectively generated around the bottom of the trap potential within the observation region. This ensures that the decay of 1*s* orthoexcitons in the observation region is not affected by inflow of excitons from outside the region.

## III. RESULTS AND DISCUSSION

We observed the time evolution of luminescence spectra of 1*s* orthoexcitons and 1*s* paraexcitons. To evaluate the conversion rate from 1*s* orthoexcitons to 1*s* paraexcitons, we concentrated on the decay in intensity of the $\Gamma_3^-$-LO-phonon-assisted luminescence from 1*s* orthoexcitons after the pulse excitation. The delay of the gate pulse with respect to the excitation pulse was systematically varied to obtain the time evolution of the luminescence intensity. To achieve adequate signal intensity with a sufficient signal-to-noise ratio, it is necessary to employ a long gate width. Nonetheless, the gate width must remain sufficiently short to accurately measure the conversion rate with high temporal resolution. Consequently, we have set the gate width to 5 ns. We varied the delay between the excitation pulse and the gate pulse in steps of 2.5 ns. We also detected the scattered excitation light from the crystal using the ICCD camera under the same delay conditions. The origin of the delay time was defined as the time when the excitation light was detected by the ICCD camera. The delay time was set from -5 ns to 100 ns. The pulse repetition frequency was 1 kHz, corresponding to the repetition frequency of the regenerative amplifier. The exposure time was typically set to 300 s. The temperature of the mixing chamber under the optical excitation was 64 mK.

Figures 3 (a-f) show the spatially resolved luminescence spectra at delay times of (a) -5 ns, (b) 0 ns, (c) 5 ns, (d) 15 ns, (e) 85 ns, and (f) 100 ns at a mixing-chamber temperature of 64 mK. The signal observed in the energy range from 2012.5 to 2014.8 meV corresponds to the $\Gamma_3^-$-LO- phonon-assisted luminescence from 1*s* orthoexcitons. The signal detected in the energy range from 2025.7 to 2028.2 meV (2017.8-2019.7 meV) is attributed to the direct recombination process of 1*s* orthoexcitons (1*s* paraexcitons). As shown in Fig. 3(b), the $\Gamma_3^-$-LO-phonon-assisted luminescence and the luminescence via the direct recombination process from 1*s* orthoexcitons appeared at a delay time of 0 ns. The spectral images in Figs. 3(c) and 3(d) at delay times of 5 and 15 ns show that the $\Gamma_3^-$-LO-phonon-assisted luminescence from 1*s* orthoexcitons decayed, while the luminescence from 1*s* paraexcitons via the direct recombination process appeared. The conversion process brings about the decay of the luminescence from 1*s* orthoexcitons and the appearance of the luminescence from 1*s* paraexcitons. At delay times of 80 and 100 ns, the luminescence from 1*s* paraexcitons was clearly observed, while the $\Gamma_3^-$-LO-phonon-assisted luminescence and the luminescence via the direct recombination process of 1*s* orthoexcitons almost disappeared. The disappearance of the signals from 1*s* orthoexcitons indicates that the conversion from 1*s* orthoexcitons to 1*s* paraexcitons was completed.

We evaluated the time evolution of the intensities of $\Gamma_3^-$-LO-phonon-assisted luminescence from 1*s* orthoexcitons and the luminescence via the direct recombination process from 1*s* orthoexcitons and 1*s* paraexcitons. We spatially integrated the signal over 100 μm around the bottom of the trap potential along the [100] crystal axis and spectrally integrated over the energy range from 2012.5 meV to 2014.8 meV to evaluate the intensity of the $\Gamma_3^-$-LO-phonon-assisted luminescence from 1*s* orthoexcitons. We also spatially integrated the signal over the same spatial region but spectrally integrated over the energy range from

†Contact author: yoshioka@fs.t.u-tokyo.ac.jp

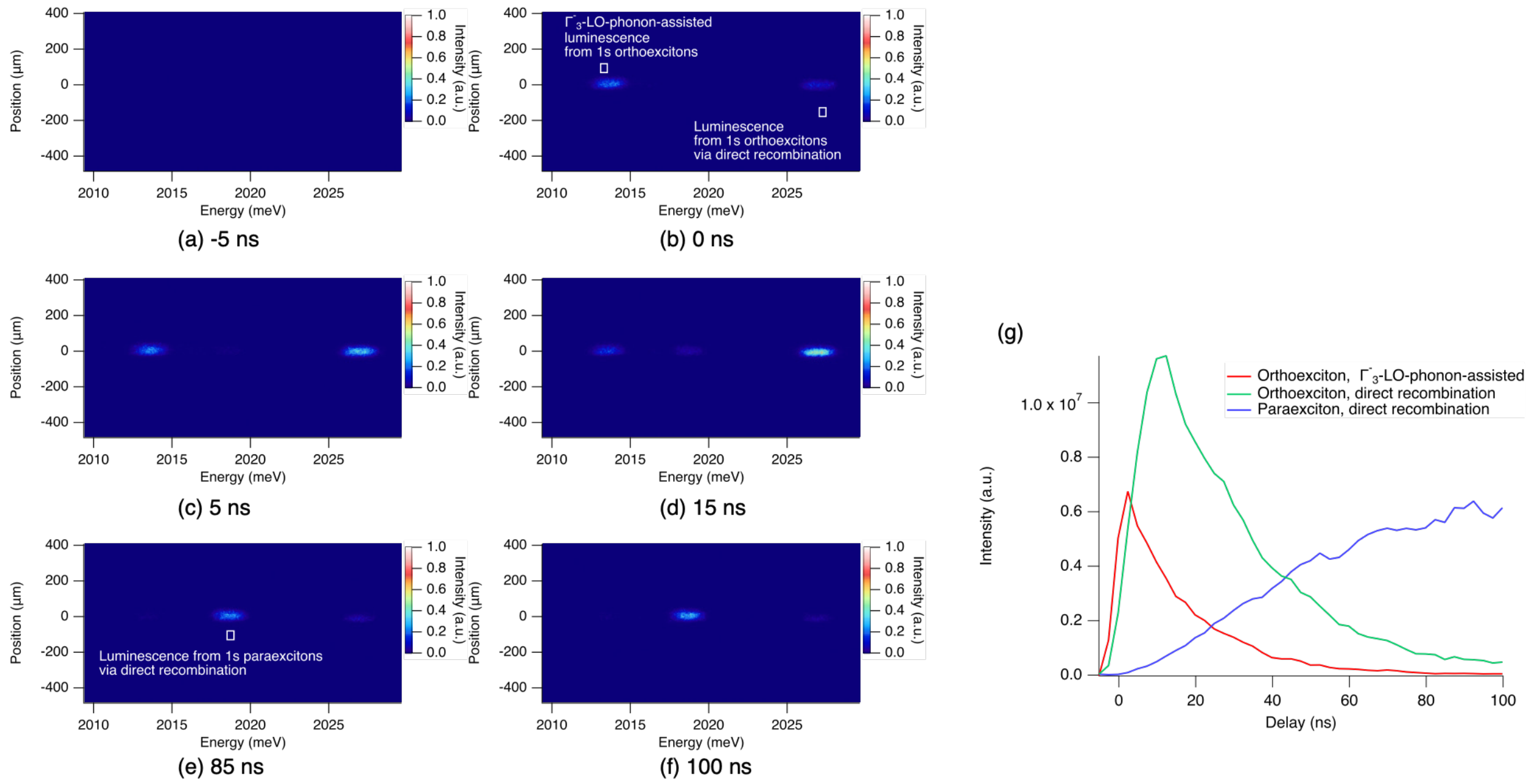

FIG. 3 (a-f) Time evolution of spatially resolved luminescence spectra of 1*s* orthoexcitons and 1*s* paraexcitons at delay times of (a) -5, (b) 0, (c) 5, (d) 15, (e) 85, (f) 100 ns. All signals in (a)–(f) were normalized to the peak intensity in (c). The signal at 2013 meV is the $\Gamma^-_3$-LO- phonon-assisted luminescence from 1*s* orthoexcitons. The signal observed at approximately 2018 meV (2026 meV) is the luminescence from 1*s* paraexcitons (1*s* orthoexcitons) via the direct recombination process. The experimental parameters were as follows: the excitation pulse energy was 300 pJ; the gate width was 5 ns; the exposure time was 150 s; the number of accumulations was 2; and the spectrometer slit width was 3.1 mm, yielding an energy resolution of 600 μeV.
(g) Time evolution of the intensity of luminescence from 1*s* orthoexcitons and 1*s* paraexcitons. The green (blue) curve shows the time evolution of the luminescence from 1*s* orthoexcitons (1*s* paraexcitons) via the direct recombination process. The red curve shows the time evolution of the $\Gamma^-_3$-LO- phonon-assisted luminescence intensity of 1*s* orthoexcitons.

2025.7 to 2028.2 meV (from 2017.8 to 2019.7 meV) to evaluate the intensity of the luminescence from 1*s* orthoexcitons (1*s* paraexcitons) via the direct recombination process.

Fig. 3(g) shows the time evolution of the two luminescence intensities from 1*s* orthoexcitons. The finite signals at negative delay times are due to the finite time resolution of the measurement. Both the luminescence via the direct recombination process and the phonon-assisted luminescence from 1*s* orthoexcitons exhibit similar exponential decay after an initial buildup of the intensity. The exponential decay is due to the decrease in the density of 1*s* orthoexcitons due to the conversion process. The intensity of the direct recombination reaches its maximum at a delay time of 10 ns, while the intensity of the phonon-assisted luminescence reaches its maximum at a delay time of 2 ns. The difference in the peak time arises from the difference in the luminescence process. As discussed in Ref. [24], in the direct recombination process that involves the emission of a single photon, the momentum conservation law requires that the translational momentum of 1*s* orthoexcitons is the same as the momentum of the emitted photon ($p_0 = 3 \times 10^{-27}$ kg m s$^{-1}$). The luminescence efficiency per unit exciton number depends on the occupation probability of excitons having $p_0$ among all exciton states. In our experiment, most 1*s* orthoexcitons have translational momenta larger than $p_0$ immediately after the generation, resulting in low luminescence efficiency. As cooling proceeds, the luminescence efficiency increases. On the other hand, in the $\Gamma^-_3$-LO-phonon-assisted luminescence, 1*s* orthoexcitons having any momentum can contribute to the luminescence. Therefore, the luminescence efficiency remains constant regardless of changes in the momentum distribution and the luminescence intensity is proportional to the number of 1*s* orthoexcitons. The difference in the time evolution of the luminescence efficiency leads to the difference in the peak time. Accordingly, time-resolved measurement of the $\Gamma^-_3$-LO-phonon-assisted luminescence is more appropriate for assessment of the conversion rate. In the present experiment, the initial kinetic energy of 1*s* orthoexcitons is expected to be limited by the spectral width of the excitation light (0.34 meV, which corresponds to approximately 4 K).

† Contact author: yoshioka@fs.t.u-tokyo.ac.jp

Our interpretation of the time evolution of the luminescence efficiency in the direct recombination process is consistent with Ref. [30]. In that study, the temperature of 1*s* orthoexcitons was reported to reach 23 K at 110 ps after pulse excitation and to gradually approach the lattice temperature at later times. Such a temperature corresponds to an average momentum of 1*s* orthoexcitons of $5 \times 10^{-26}$ kg m $s^{-1}$, which is 15 times larger than $p_0$. Thus, most 1s orthoexcitons at 23 K occupy momentum states larger than $p_0$, resulting in low luminescence efficiency. The subsequent cooling of 1*s* orthoexcitons toward the lattice temperature is consistent with the increase in the direct-recombination luminescence efficiency discussed above.

Fig. 3(g) also shows that the intensity of the luminescence from 1*s* paraexcitons increases almost monotonically from a delay time of 0 ns to 100 ns. The conversion from 1*s* orthoexcitons to 1*s* paraexcitons contributes to this monotonic increase. However, the intensity of the $\Gamma^-_3$-LO- phonon-assisted luminescence from 1*s* orthoexcitons at a delay time of 64 ns is only 3% of the peak intensity. This suggests that the conversion process is almost completed at the delay time and the increase in the density of paraexcitons stops at delay times greater than 64 ns. Nevertheless, the intensity of the luminescence from 1*s* paraexcitons continues to increase almost monotonically at those delay times. As discussed in Ref. [23], the temperature of paraexcitons gradually decreases to 270 mK over 90 ns when the mixing-chamber temperature is below 100 mK. This gradual temperature decrease results in an increase in the luminescence efficiency of the direct recombination. The increase in the luminescence efficiency leads to the monotonic increase at those delay times.

We extracted the conversion rate from the time evolution of the intensity of the $\Gamma^-_3$-LO-phonon-assisted luminescence at a mixing-chamber temperature of 64 mK. The intensity of the phonon-assisted luminescence corresponds to the number of 1*s* orthoexcitons in the observation region. The rate equation for the number of 1*s* orthoexcitons is given as follows:

$$\frac{dN_o(t)}{dt} = -\frac{N_o(t)}{\tau} + g(t), \qquad (1)$$

where $N_o(t)$, $1/\tau$, and $g(t)$ denote the number of 1*s* orthoexcitons in the observation region as a function of time, the conversion rate, and generation rate of 1*s* orthoexcitons as a function of time, respectively. As mentioned above, the inflow of 1*s* orthoexcitons from

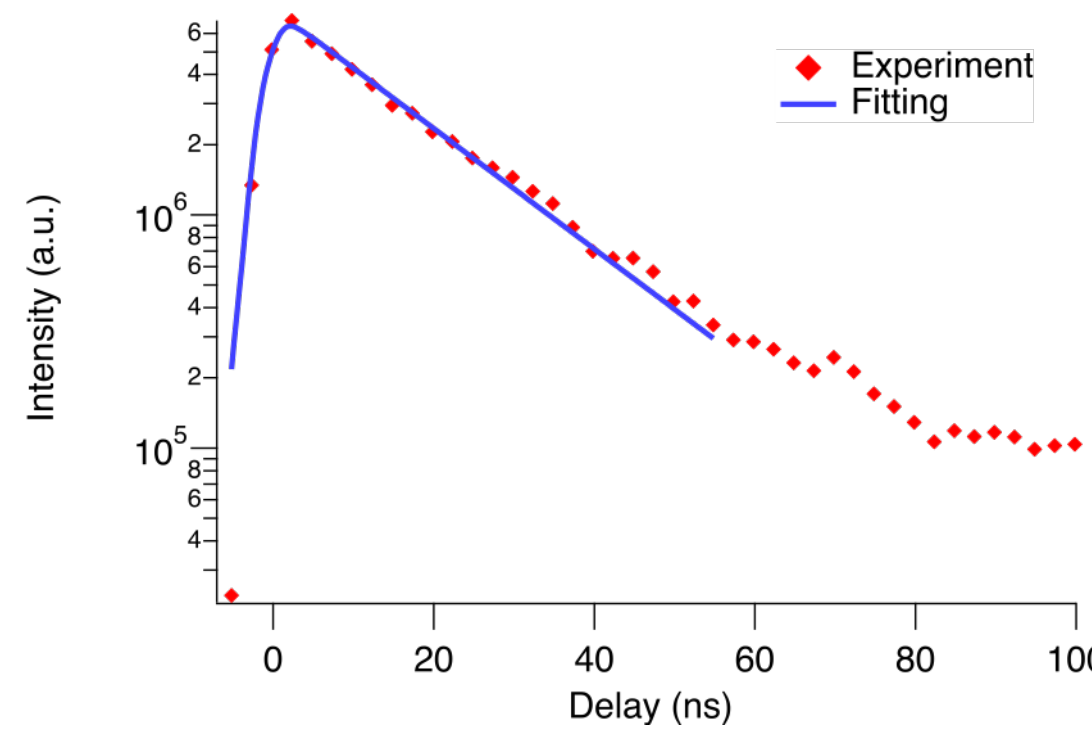

FIG. 4. Time evolution of the intensity of the $\Gamma^-_3$-LO- phonon-assisted luminescence of 1*s* orthoexcitons. The red diamonds represent the luminescence intensity as a function of time in the range from -5 to 100 ns. The blue curve shows the calculated time evolution of the luminescence intensity as a function of time in the range from -5 to 55 ns.

outside the observation region did not occur because 1*s* orthoexcitons were generated only in the observation region. Therefore, the rate equation includes only the decay term and the generation term. We used the excitation pulse train with a repetition rate of 1 kHz and a pulse duration of approximately 100 picoseconds. The pulse period is much longer than the lifetimes of 1*s* orthoexcitons and 1*s* paraexcitons. Even though the pulse train successively excites the crystal, the buildup of 1*s* orthoexcitons generated by preceding pulses can be neglected. Therefore, it is adequate to focus on the density dynamics of 1*s* orthoexcitons induced by a single pulse. The pulse duration of the excitation light was shorter than the gate width of 5 ns. 1*s* orthoexcitons are generated instantaneously when the absorption of photons in the crystal occurs. Therefore, we model $g(t)$ as a delta function:

$$g(t) = N_1 \delta(t - t_0), \qquad (2)$$

where $N_1$ and $t_0$ denote the number of 1*s* orthoexcitons generated per one pulse and the time when the excitation pulse enters the crystal, respectively. As a result, we obtain the following equation using Eq. (1) and (2):

For $t < t_0$,

$$N_o = 0\,. \qquad (3)$$

For $t_0 \leq t$,

$$N_o = N_1 \exp(-\frac{t-t_0}{\tau}). \qquad (4)$$

Taking into account the temporal resolution, we fit the time-resolved luminescence intensity $I_o(t)$ with the following expression:

$$I_o(t) = \int_{-\infty}^{\infty} du\, N_o(t-u) \cdot h(u)\,, \qquad (5)$$

†Contact author: yoshioka@fs.t.u-tokyo.ac.jp

where $h(t)$ is the instrument response function, which we obtained from the time-resolved measurement of the scattered excitation light under the same gate conditions. Because the response function $h(t)$ in Eq. (5) is measured with the same excitation pulse, the finite pulse duration is included in $h(t)$. The fitting parameters were the conversion rate $1/\tau$, the amplitude, and $t_0$. The fitting was restricted to delay times less than 56 ns because the decay of the luminescence intensity deviates from an exponential decay at delay times exceeding 56 ns. A possible origin is the decrease in the conversion rate as the orthoexcitons cool, as discussed below. As shown in Fig. 4, the calculated result reproduces the experimental time evolution of the luminescence intensity well. From our fitting, we estimated the conversion rate to be $(5.9 \pm 0.1) \times 10^7$ $s^{-1}$, which corresponds to a conversion time of 17 ns, at the mixing-chamber temperature of 64 mK. This value is significantly smaller than the value of $1.6\times 10^8$ $s^{-1}$ at 2 K under an applied stress of 2.2 kbar, which corresponds to the same stress condition as in our experiment, reported in Ref. [19].

In addition, we measured the conversion rate as a function of the mixing-chamber temperature, which varied from 64 mK to 800 mK. The temperature range was constrained by the dilution refrigerator, whose cooling mechanism works only below 870 mK. The estimation of the conversion rate was based on time-resolved measurements of the $\Gamma^-_3$-LO-phonon-assisted luminescence intensity of 1*s* orthoexcitons, as explained above. Fig. 5 shows that the conversion rate slightly increases from $5.9 \times 10^7$ $s^{-1}$ to $6.3 \times 10^7$ $s^{-1}$ as the mixing-chamber temperature is increased from 64 mK to 800 mK. As discussed in Sec. I, the conversion rate has conventionally been assumed to be temperature-independent below 2 K, taking a constant value determined by the applied stress. Under this assumption, the conversion rate at an applied stress of 2.2 kbar would be constant at $1.6 \times 10^8$ $s^{-1}$ below 2 K. However, the conversion rates we measured at sub-Kelvin temperatures were approximately half the expected value, indicating a clear deviation from the conventional prediction.

We attribute this deviation to the difference between the orthoexciton temperature and the lattice temperature in the previous measurements below about 3 K, as discussed qualitatively below. Fig. 1 shows the temperature dependence of the conversion rate, including both previously reported data and our results. Following conventional practice, we plot our data as a function of the lattice temperature, which implicitly assumes that the orthoexciton temperature is equal to the lattice temperature. The validity of this assumption, however, depends on the lattice-temperature range. At lattice temperatures above 3 K, the thermal relaxation time between orthoexcitons and the lattice via LA-phonon emission is less than a nanosecond [31]. This relaxation time is well below the conversion time of about 3 ns at a lattice temperature of 3 K [19]. The orthoexcitons therefore are thermalized with the lattice for most of the conversion time. At lattice temperatures below 3 K, however, thermal relaxation via LA-phonon emission is suppressed [26, 30, 31]. This suppression results from the freeze-out of LA- phonon emission for orthoexcitons with translational momenta below $1.1 \times 10^{-26}$ kg m $s^{-1}$. This translational momentum corresponds to a kinetic energy equivalent to 1.8 K Consequently, approximately 80% of the orthoexciton population cannot emit LA phonon, even at an orthoexciton temperature of 3 K. The orthoexciton temperature therefore does not decrease appreciably as the lattice temperature is reduced below 3 K. Therefore, the previous measurements at different lattice temperatures below 3 K may have been performed at nearly the same orthoexciton temperature, which would account for the nearly constant conversion rates reported in this range. Note that, even at lattice temperatures below 3 K, thermal relaxation between orthoexcitons and the lattice via TA-phonon emission is enhanced under applied stress. This enhanced thermal relaxation allows orthoexcitons to cool below 3 K. In our measurements, orthoexcitons were generated under an applied stress of 2.2 kbar at sub-Kelvin temperatures. Accordingly, our conversion rates were obtained at orthoexciton temperatures well below those in previous studies. The difference in the conversion rate between our results and those of previous studies is thus likely to originate from the difference in the orthoexciton temperature, which is manifested as an apparent deviation when the conversion rate is plotted as a function of the lattice temperature. It remains to be investigated whether our results and previous reports connect smoothly when plotted as a function of the orthoexciton temperature. Note that the reference value at 2 K was also measured under stress [19]. In that case, the orthoexcitons can cool to approximately 2 K by TA-phonon emission, and the difference from our results reflects the further decrease of the orthoexciton temperature below 2 K.

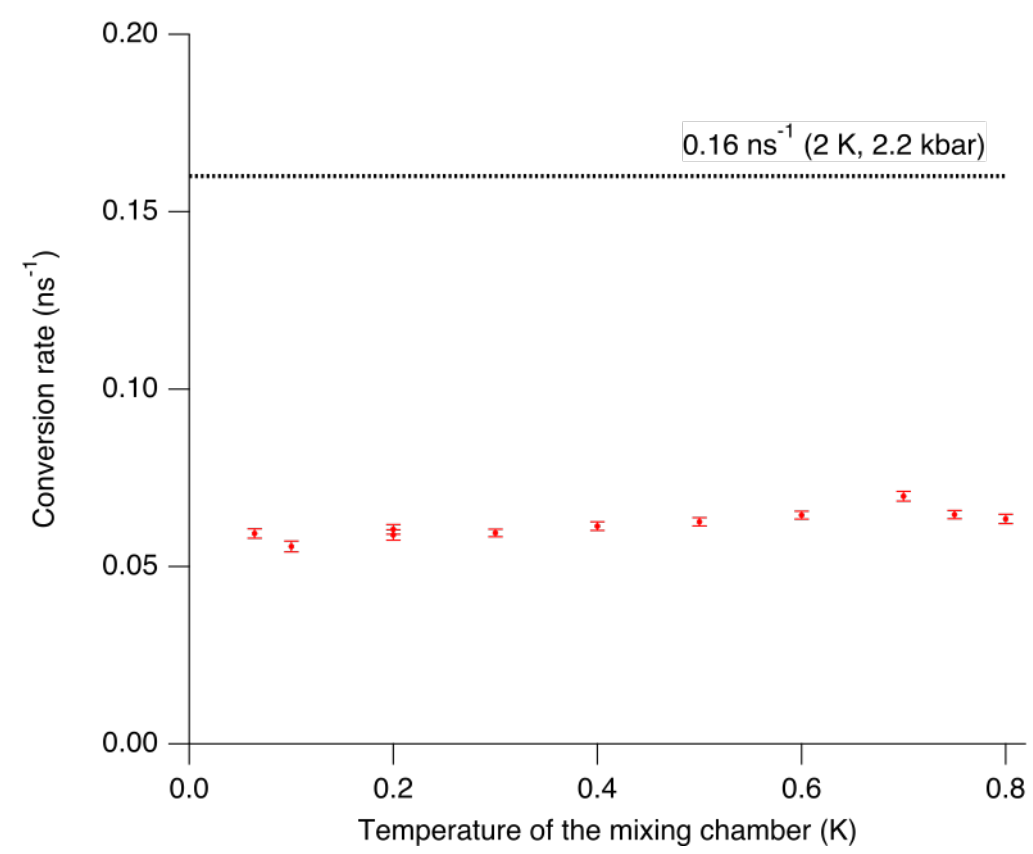

FIG. 5. Ortho-para conversion rate as a function of the mixing-chamber temperature. The red dots show the evaluated conversion rates at mixing-chamber temperatures of 64, 100, 200, 300, 400, 500, 600, 700, 760, and 800 mK. Error bars show the standard errors of the fit. The dotted line shows the conversion rate of $1.6 \times 10^8$ $s^{-1}$ [19] at 2 K under 2.2 kbar, which is the same stress condition as in our experiment.

† Contact author: yoshioka@fs.t.u-tokyo.ac.jp

A deeper understanding of the conversion process at sub-Kelvin temperatures requires not only determining the effective temperature of orthoexcitons but also characterizing the momentum distribution of orthoexcitons and the momentum dependence of the conversion rate. Although the enhanced TA-phonon emission under applied stress allows orthoexcitons to cool below 3 K, this cooling process may take about 100 ns to cool down to a few hundred millikelvin, judging from the thermal relaxation time reported for paraexcitons [23]. The relaxation time is roughly ten times longer than the conversion timescale reported here, which is about 17 ns at the lattice temperature of 64 mK. Consequently, the momentum distribution of orthoexcitons cannot be described by a thermal distribution with a well-defined effective temperature. In our study, the conversion rate was measured at various lattice temperatures below 1 K. The momentum distribution of orthoexcitons during the conversion process at sub-Kelvin temperatures is governed primarily by TA-phonon emission. Phonon absorption is weak compared with phonon emission at sub-Kelvin temperatures, where the occupation of the relevant phonon modes is small. As a result, the momentum distribution depends only weakly on the lattice temperature. Although the momentum dependence of the conversion rate remains to be studied, this weak dependence may explain the weak lattice-temperature dependence of the conversion rates that we observed.

Future work should address the thermal relaxation dynamics of orthoexcitons below 1 K during the conversion process, as well as the momentum dependence of the conversion rate. These investigations will help resolve the discrepancy in the conversion rate below 2 K revealed in this work. More broadly, they call for further examination of the microscopic mechanism of the orthoexciton-paraexciton conversion, so that a single theoretical framework can describe the conversion over the entire range of lattice temperatures, from sub-Kelvin to well above 2 K.

## IV. CONCLUSIONS

We evaluated the rate of the conversion from 1*s* orthoexcitons to 1*s* paraexcitons at mixing-chamber temperatures below 1 K. To determine the conversion rate, we measured the time evolution of the $\Gamma^-_3$-LO-phonon-assisted luminescence intensity of 1*s* orthoexcitons generated by pulsed excitation. We obtained a conversion rate of $(5.9 \pm 0.1) \times 10^7 s^{-1}$ at a mixing-chamber temperature of 64 mK under an applied stress of 2.2 kbar. This value is significantly smaller than the previously reported value of $1.6 \times 10^8$ $s^{-1}$ at a lattice temperature of 2 K under the same applied stress. We also measured the lattice temperature dependence of the conversion rate over the range from 64 mK to 800 mK. The measured conversion rates were less than half of the value reported at 2 K and depended only weakly on the lattice temperature. These results are inconsistent with the conventional expectation that the conversion rate remains constant below 2 K.

This deviation indicates that the conventional expectation of a constant conversion rate below 2 K is not valid at least under applied stress. At lattice temperatures below 3 K, the thermal relaxation of orthoexcitons becomes comparable to, or even slower than, the conversion, in contrast to the situation at higher temperatures. As a result, the orthoexciton temperature no longer follows the lattice temperature closely, which can give the appearance of a conversion rate converging to a constant value at low temperatures. In reality, the orthoexciton temperature may continue to decrease in the sub-Kelvin regime, although more gradually than at lattice temperatures above 2 K. In this regime, where the evolution of a non-thermal distribution and the ortho-to-para conversion proceed on comparable timescales, we have argued that the conversion rate should be understood in terms of the orthoexciton momentum distribution rather than the lattice temperature. Future measurements that probe both the momentum-space dynamics of orthoexcitons and the ortho-to-para conversion are expected to test the interpretation presented above. Such measurements would also provide a basis for revisiting the conversion mechanism, because the conventional models have not been tested below a lattice temperature of 2 K.

Previous studies have also evaluated the stress dependence of the conversion rate at a fixed lattice temperature of 2 K, and any revised conversion mechanism must reproduce this dependence as well. Measurements of the stress dependence in the sub-

†Contact author: yoshioka@fs.t.u-tokyo.ac.jp

Kelvin regime will be equally important for testing the revised theory.

In the present study, we developed space- and time-resolved luminescence spectroscopy with nanosecond time resolution at sub-Kelvin temperatures. This technique provides an experimental basis for elucidating the formation dynamics of an exciton Bose-Einstein condensate. Reference [24] reported the direct observation of a condensate of 1*s* paraexciton below 400 mK. The study found an unconventionally small condensate fraction of 0.016, which is two orders of magnitude smaller than that expected for an ideal Bose gas. Understanding the formation dynamics of the paraexciton condensate is essential for clarifying the origin of this unusually small condensate fraction. At an earlier stage of condensate formation, 1*s* paraexcitons are generated through the conversion process at a high density of about $10^{15}$ $cm^{-3}$ [24]. Our measurement of the conversion rate provides direct information on the density dynamics of the 1*s* paraexcitons during this initial stage. Furthermore, the present space- and time-resolved spectroscopy can be extended to investigate the thermal relaxation dynamics of 1*s* paraexcitons below 1 K. These investigations will pave the way toward essential understanding of the formation dynamics.

Rydberg excitons have recently attracted considerable interests due to potential applications in quantum information techniques. In particular, $Cu_2O$ is one of the most well-known materials for studying Rydberg excitons. The coherence time is an essential parameter for any quantum technology, and the lifetime is a fundamental source of decoherence [32]. Therefore, our measurement of the rate of the spin conversion process, which may predominantly limit the lifetime, provides a basis of study of coherence time of Rydberg excitons in $Cu_2O$.

## ACKNOWLEDGMENTS

We thank K. Saito for assistance with data taking and analysis, and M. Kuwata-Gonokami for discussions. This research was supported by MEXT, JSPS KAKENHI (Grant Nos. JP26247049, JP17H06205, JP21H04673, JP24H00217), by the MEXT Quantum Leap Flagship Program (MEXT Q-LEAP), Grant Number JPMXS0118067246, and by the RIKEN TRIP initiative (HIKARI-COOL Tokyo).

† Contact author: yoshioka@fs.t.u-tokyo.ac.jp

† Contact author: yoshioka@fs.t.u-tokyo.ac.jp